# Unraveling Momentum and Heat Intercoupling in Reattaching Turbulent Boundary Layers Using Dynamic Mode Decomposition

Rozie Zangeneh[1]
[1]*Department of Mechanical Engineering, Stanford University, Stanford, CA, USA ,*

*Email address: rozieza@stanford.edu*

Dynamic mode decomposition method is deployed to investigate the heat transfer mechanism in a compressible turbulent shear layer and shockwave. To this end, highly resolved Large Eddy Simulations are performed to explore the effect of wall thermal conditions on the behavior of a reattaching free shear layer interacting with an oblique shock in compressible turbulent flows. Various wall temperature conditions, such as cold adiabatic and hot wall, are considered. Dynamic mode decomposition is used to isolate and study the structures generated by the shear layer exposed in the boundary layer. Results reveal that the shear layer flapping is the most energetic mode. The hot wall gains the highest amplitude for the flapping frequency, and the vortical motions are most intense in the vicinity of the reattachment point of the heated wall. The vortex shedding due to the large-scale motion of the shear layer is associated with the second energetic mode. The cold wall not only has a higher amplitude of the shedding mode, but it also has a lower frequency compared to the adiabatic and hot walls. This work sheds light on the underlying physics of the nonlinear intercoupling between momentum and heat, thereby enhancing high-speed flight efficiency and reducing aircraft fatigue from intense wall-pressure fluctuations and heat flux.

## 1 INTRODUCTION

The phenomenon of flow separation and reattachment is encountered throughout the field of fluid flows and often results in diminishing the aerodynamic performance. Since the separation of flow may be used for controlling the heat transfer to a surface and for aerodynamic control purposes, the heat transfer characteristics of separated flows are of considerable interest in designing high-speed vehicles [1]. In high-speed flows, the problem of separation is often a multi-physics problem that involves the effects of compressibility, shockwave, and boundary layer interaction, as well as aerodynamic heating. Extensive studies have been conducted over the past few decades in order to understand the mechanism of flow unsteadiness in the separated turbulent boundary layer (TBL) in view of mitigating aircraft fatigue loading caused by the intense wall pressure fluctuations and heat flux accompanied by shock unsteadiness. Perhaps the first study focusing primarily on this problem was carried out by Kistler [2], who measured wall pressure fluctuations. His results showed an intermittent wall pressure signal in a supersonic, forward-facing step. Further experimental data on turbulent separated flows with heat transfer involving steps, spoilers, and cavities have been obtained by Gadd [3], Larson [4], Thomman [5], and Charwat et al. [6]. Lighthill [7] has described the separation of subsonic and supersonic flow in relation to the upstream influence of disturbances through the boundary layer.

Hadjadj [8] investigated oscillations of the reflected shock occurring at low frequencies and concluded that the instabilities of the reflected shock are due to the intrinsic low-frequency movement of the dynamically coupled shock and separation bubble. The shock motion has two characteristic scales: one because of perturbation of the shock by organized structures in the flow turbulence, and the other because of relatively large-scale, low-frequency expansion and contraction of the separation bubble (Poggie and Smits, 1997 [9]). The shock motion may be driven in part by vortex shedding in low speed separated flows (Eaton et al., [10]; Driver et al., [11]) and also, by turbulence in the incoming flow and the separation bubble (Bogar, [12]; Brusniak and Dolling, [13]). Profiles of the intensity of pressure fluctuations and heat transfer fluctuations at the wall of a separated boundary layer in compressible flow have been shown to have two distinct maxima that are an order of magnitude above the fluctuation levels in the incoming TBL (Dolling and Murphy, [14]; Hayashi et al., [15]). Poggie and Smits [9] tested the influence of disturbances in the incoming shear layer through steady air injection in the vicinity of separation and observed a substantial increase in the amplitude of the shock oscillation and also a distinct shift to a lower frequency in the power spectra of the pressure fluctuations. Dussauge et al. [16] evaluated unsteadiness in shockwave and TBL interactions with separation and observed that the fluctuations produced by the shock motion are much lower than the characteristic frequencies of turbulence in the incoming boundary layers.

Most prior research has been expended around the case of adiabatic wall conditions on shock and TBL interactions. Effort has been invested in the last decade to characterize the large-scale, low-frequency unsteadiness, which is typical of the interaction region. The influence of wall thermal conditions on the characteristics of shock-TBL interaction is found to be considerable, and wall cooling is often proposed as a potential flow control technique. It was reported that strong cooling is capable of delaying the laminar-turbulent boundary layer transition toward higher Reynolds numbers as well as reducing the thickness of the subsonic layer by changing the local speed of sound [17]. In this context, the effects of heat transfer in turbulent interactions over a compression ramp have been investigated by Spaid and Frishett [18], who performed experiments in supersonic regimes by considering a cold wall and an adiabatic wall. The measured data showed that decreasing the wall temperature increases the incipient separation angle and decreases the separation distance. Hayashi and Sakurai [19] reported the measurement of heat transfer by considering a supersonic TBL developing over an isothermal cold wall interacting with an oblique shock at various incident angles. They observed a complex spatial variation of the heat transfer coefficient, characterized by a rapid increase near the separation point, followed by a sharp reduction within the separation bubble and a further increase in the reattachment point's proximity. Recently, measurements of skin friction and heat transfer have been reported by Schülein [20], who considered an impinging shock at various incident angles.

In recent years, considerable progress has been achieved in predicting reattaching free shear layer for compressible turbulent flows using Direct Numerical Simulation (DNS) and Large Eddy Simulation (LES) [21]. However, DNS or LES results for surface heat transfer in shock and TBL interactions are rare. Since Knight et al. [22] have raised attention to the fact that insufficient DNS and LES results for the surface heat transfer in shockwave and TBL interaction were available, only a few high-fidelity simulations have been carried out to explore the effect of either wall heating or cooling on the flow unsteadiness. Bernardini et al. [23] conducted DNS to investigate the impact of wall thermal conditions on the behavior of oblique shockwave-TBL interactions at a freestream Mach number of 2.28. Their results showed that the main effect of cooling is to decrease the characteristic scales of the interaction in terms of upstream influence and the extent of the separation bubble. Volpiani et al. [24] conducted DNS for a similar case and found that the thermal boundary condition at the wall has a large effect on the size of the interaction region and the level of pressure fluctuations. Their results showed a good agreement with experimental studies and confirmed the substantial heat transfer and intricate pattern in the interaction region.

In a recent study by Zangeneh [25], it was shown that wall heating moves the reattachment location downstream. On the contrary, wall cooling moves it upstream on the ramp but has the drawback of increasing the wall pressure fluctuations. Results show that downstream of the mean reattachment point, the skin friction coefficient is maximized by wall cooling and minimized by wall heating. For the cooled wall case, the growth of thermal structures is observed, and the local Stanton number exhibits a strong intermittent behavior in the interaction region with intense heat transfer rates. The effect of wall cooling on the intensity of the wall pressure fluctuation was also shown in the recent work of Hirai and Kawai [26].

The physics mentioned above suggests strong coupling between momentum and heat in a reattaching shear layer; therefore, the dynamics of the reattaching shear layer are strongly affected by the wall's thermal condition. In this study, we use the dynamic mode decomposition (DMD) method to isolate and study the structures generated by the shear layer exposed in the boundary layer. Dynamic mode decomposition (DMD) is a data-driven computational technique capable of extracting a linear system from flowfields measured in physical experiments or generated by simulation ( [27] ), which extracts the most energetic motions of the flowfield. To date, no mode decomposition method has been applied to the system of separated and reattached flows with heat transfer to explain the underlying physics of the nonlinear intercoupling between heat and momentum. Hence, this work sheds light on the underlying mechanism of the intercoupling of momentum and heat and provides guidelines for controlling high-speed flight vehicles and mitigating aircraft fatigue loading caused by intense wall pressure fluctuations and heat flux.

The paper is organized as follows: Section 2 provides details on the methodology, and the DMD method is described in Section 3. Section 4 presents the results of simulations for three wall temperature conditions with DMD applied to simulations and finally summarizes the main findings of this work.

# 2 METHODOLOGY

## 2.1 Flow Model and Computational Arrangement

**Table 1. Flow parameters for LES simulations.**

| **Test case** | $M_\infty$ | $T_w/T_r$ | $T_w/T_\infty$ | $\delta_s(mm)$ | Line |
|---|---|---|---|---|---|
| Cooled | 2.92 | 0.5 | 1.35 | 3 | — |
| Adiabatic | 2.92 | 1 | 2.69 | 3 | — |
| Heated | 2.92 | 1.5 | 4.11 | 3 | — |

Figure 1 illustrates a schematic view of the domain configuration for the reattaching shear layer. The configuration was originally designed by Settles et al. [28], and basic properties of the flow field were investigated in numerous experimental campaigns [19, 21]. An equilibrium TBL develops on a plate. The layer then separates at a sharp corner and bridges a 2.54 cm deep cavity. It reattaches on a ramp inclined 20° to the horizontal, interacting with an oblique shock wave. In the simulations, the freestream temperature is $100K$ with the stagnation temperature of $290K$. The freestream Mach number is 2.92 at $\frac{\rho_\infty u_\infty}{\mu_\infty} = 6.7 \times 10^7$, where $\rho_\infty$ is the freestream density, $u_\infty$ is the freestream velocity, and $\mu_\infty$ is the freestream viscosity. The viscosity depends on the temperature by the Sutherland law, $\frac{\mu}{\mu_\infty} = (\frac{T}{T_\infty})^{1.5}\frac{T_\infty+T_s}{T+T_s}$, where $T_\infty$ and $T_s$ are the freestream temperature and the Sutherland constant, respectively [29]. The zero-pressure-gradient equilibrium TBL in the vicinity of the backward-facing step has a thickness of about $\delta_s = 3mm$ and a momentum thickness Reynolds number of about $R_\theta = 10^4$, evaluated at the vicinity of the backward-facing step ($\delta_s = 3mm$). Three simulations have been carried out at various values of the wall-to-recovery-temperature ratio on the ramp, representing cooled ($T_w/Tr = 0.5$), adiabatic ($T_w/T_r = 1.0$), and heated ($T_w/T_r = 1.5$) walls as shown in Table 1, where $T_r$ is the wall recovery temperature.

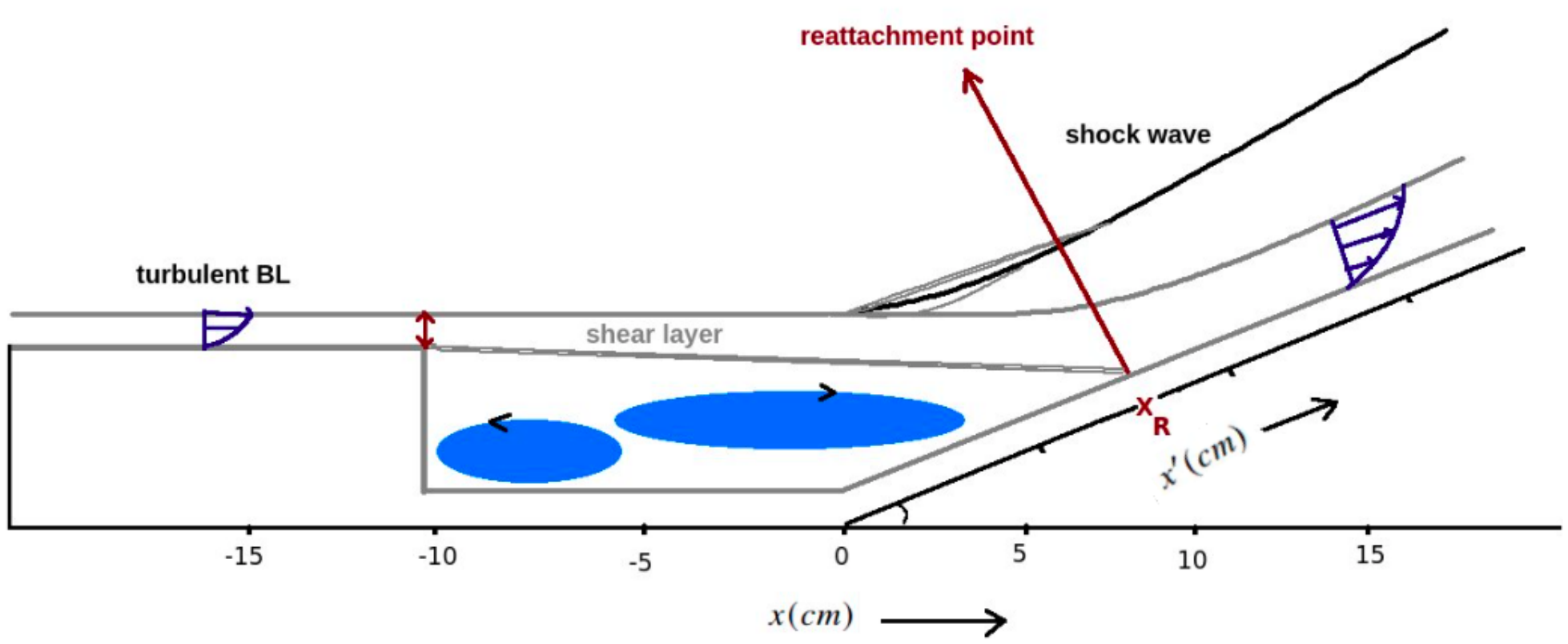

**Figure 1. Mean flow field for reattaching shear layer in a scramjet. Recreated from Settles et al. [28].**

## 2.2 Grid Convergence

The computational domain is depicted in Fig. 2. The plate is extended to achieve the desired boundary layer thickness at the corner. The choice of the grid resolution is based on resolving the relevant scales in the boundary layer. The computational domain is split into three regions for meshing: the upstream region (I), the cavity region (II), and the ramp (III). Section I, II, and III consist of $421 \times 111 \times 321$, $180 \times 161 \times 321$, and $950 \times 161 \times 321$ grids respectively as illustrated in Fig 2. The choice of the grid resolution is based on resolving the relevant scales in the boundary layer. The grid size expands in the wall-normal direction in order to cluster grid points near the wall; therefore, the grid size changes from $\Delta y^+ \leq 1$ adjacent to the wall to $\Delta y^+ = 13$ at the end of the domain. Streamwise clustering is performed near the step to resolve the vortex roll-up due to Kelvin–Helmholtz instability, and near the ramp to accurately capture the flow instability near the reattachment point. Additionally, the grids are clustered in the shear layer region. The streamwise and spanwise spacings are $\Delta x^+ \leq 8.1$ and $\Delta z^+ = 7.5$, respectively. Note that this grid spacing is smaller than the mesh size usually employed for LES of shock-TBL interaction with adiabatic walls, due to the reduction in the viscous length scale when cooling is applied to the wall. To evaluate convergence as the mesh size increases, three computations are conducted on a sequence of three meshes with grid resolutions Coarse, Medium, and Fine. In the LES approach, the boundary layer thickness dictates the grid resolution requirement in the spanwise direction [30], so the number of cells in this direction is fixed. A constant refinement ratio, $r = 1.3$, is applied between the three grid resolutions in the streamwise and wall-normal directions. Figure 3 shows the convergence results for the skin friction coefficient, $C_f = 2\tau_w/\rho_\infty u_\infty^2$ [25], of the cold wall case. The numerical solution converges to a unique value of $C_f$ as the grid density increases. It is apparent that the coarse grid does not resolve the separation since the $C_f$ values are near zero in the separation region. In contrast, the medium and fine meshes resolve the separation bubble and tend to converge. The medium mesh is selected for the simulations, and the grid details are provided above.

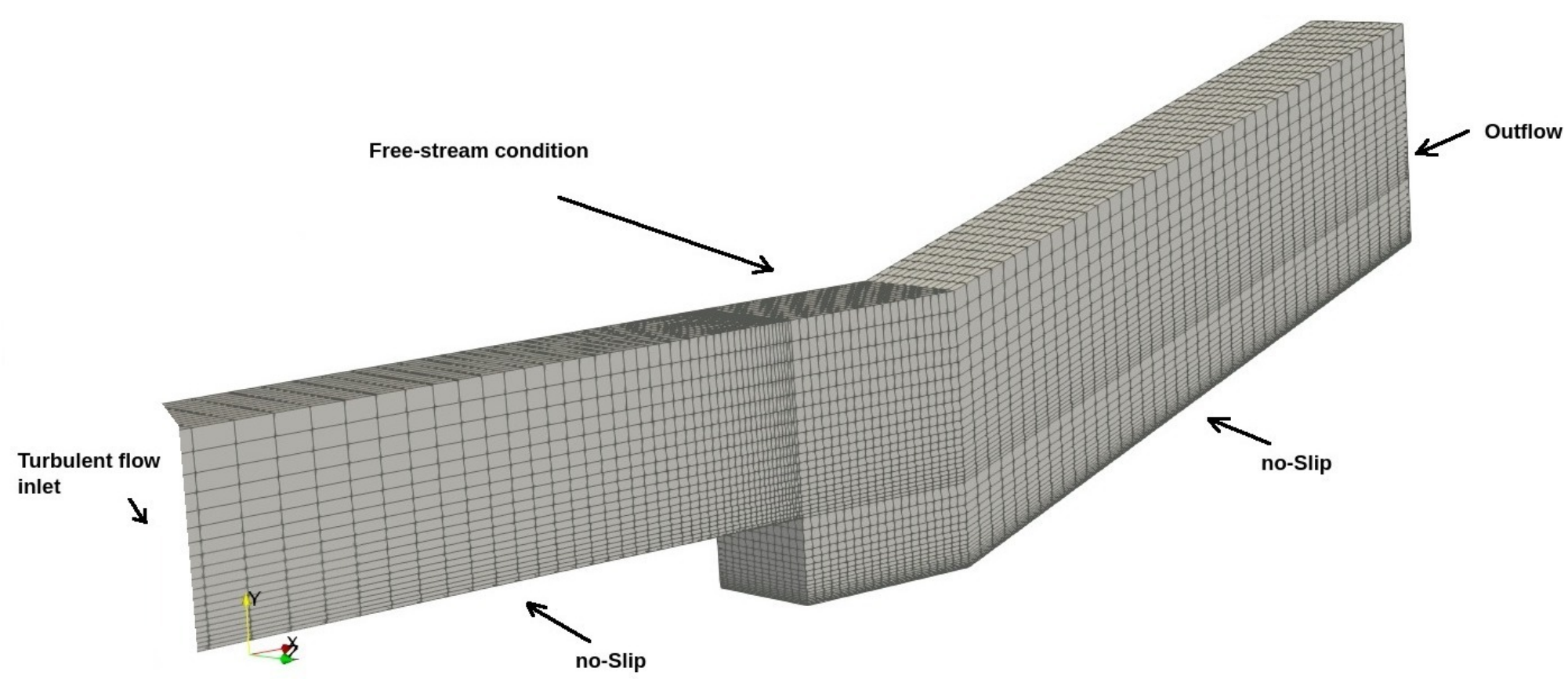

**Figure 2. Grid and boundary conditions for the simulation domain (Mesh is coarsened for visual clarity).**

No-slip condition is applied to the walls, and periodical boundary conditions are applied in the spanwise direction. A digital-filter-based inlet boundary condition is used to generate synthetic turbulence-like time series. The digital-filter approach for generating artificial inflow data is to produce a velocity signal with specific statistical properties, which may, for example, be known from experimental data. Such quantities are mean values, fluctuations, higher-order moments, length, and time scales. For more details of the theoretical expressions of this method, refer to the original work of Klein et al. [31]. The choice of inflow development length, $L_{dev}$, is made by some trial-and-error to obtain a fully developed turbulence state with the desired boundary layer thickness, $\delta_s$, at the step. A non-reflecting boundary condition is imposed to attempt to reconstruct this kind of non-reflective scheme to allow the sound wave to flow smoothly out of the domain through the outlet. This boundary condition was proposed by Poinsot et al. [32], which attempts to reconstruct a non-reflecting scheme to avoid spurious numerical occurrences near the exit.

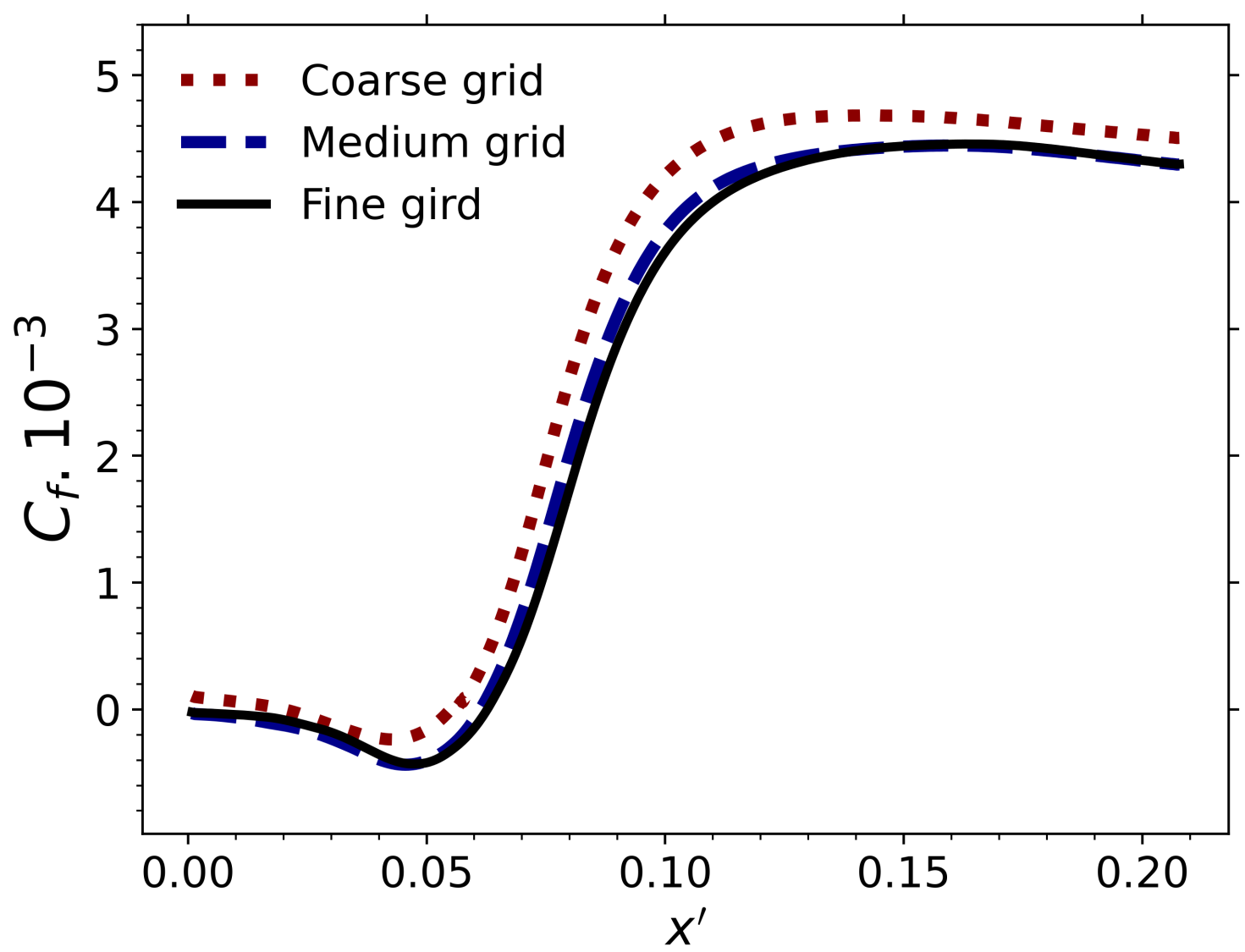

**Figure 3. Comparison of three grid resolutions for computing $C_f$ for the cold wall case.**

## 2.3 Governing Equations

For LES simulation, the filleted compressible Navier-Stokes equations are solved [33] as follows:

$$\frac{\partial \bar{\rho}}{\partial t} + \frac{\partial (\bar{\rho}\tilde{u}_j)}{\partial x_j} = 0, \tag{1}$$

$$\frac{\partial (\bar{\rho}\tilde{u}_i)}{\partial t} + \frac{\partial (\bar{\rho}\tilde{u}_i\tilde{u}_j)}{\partial x_j} = -\frac{\partial \bar{p}}{\partial x_i} + \frac{\partial 2\mu\tilde{S}_{ij}}{\partial x_j} - \frac{\partial \tau_{ij}}{\partial x_j}, \tag{2}$$

$$\frac{\partial (\bar{\rho}\tilde{e})}{\partial t} + \frac{\partial}{\partial x_j}\left(\bar{\rho}\tilde{e}\tilde{u}_j + \bar{p}\tilde{u}_j\right) = \frac{\partial}{\partial x_j}\left((2\mu\tilde{S}_{ij} - \tau_{ij})\tilde{u}_i\right) + \frac{\partial}{\partial x_j}\left(\lambda\frac{\partial \tilde{T}}{\partial x_j}\right) - \frac{\partial q_j}{\partial x_j}, \tag{3}$$

where $\tau_{ij} = \bar{\rho}\widetilde{u_i u_j} - \bar{\rho}\tilde{u}_i\tilde{u}_j$ is the subgrid-scale (SGS) stress tensor, and $q_j = \bar{\rho}C_p\widetilde{u_j T} - \bar{\rho}C_p\tilde{u}_j\tilde{T}$, the SGS heat flux, representing the effect of the small scales of turbulence on the large ones. Here ( $\bar{}$ ) and ( $\tilde{}$ ) imply the averaged value and the density-weighted averaged (Favre-averaged) value, respectively. The set of equations is closed by setting $\tau_{ij} = -2\mu_t\widetilde{S_{ij}}$ and $q_j = -C_p\frac{\mu_t}{Pr_t}\frac{\partial \tilde{T}}{\partial x_j}$ where $\widetilde{S_{ij}} \equiv \frac{1}{2}\left(\frac{\partial \widetilde{u_i}}{\partial x_j} + \frac{\partial \widetilde{u_j}}{\partial x_i}\right) - \frac{1}{3}\frac{\partial \widetilde{u_k}}{\partial x_k}\delta_{ij}$ and $\mu_t$ is the eddy viscosity, and by the modified averaged Equation of State, $\bar{P} = \bar{\rho}R\tilde{T}$. k-Equation model is used to model the eddy viscosity, $\mu_t$. The turbulent viscosity is given by: [34]:

$$\mu_t = \bar{\rho}C_k\Delta k^{0.5}. \tag{4}$$

The transport equation for kinetic energy is solved to account for the effects of convection, diffusion, production, and destruction on the subgrid tensor as follows:

$$\frac{\partial (\bar{\rho}k)}{\partial t} + \frac{\partial (\bar{\rho}\tilde{u}_j k)}{\partial x_j} - \frac{\partial}{\partial x_j}\left[\bar{\rho}\left(\nu + \nu_t\right)\frac{\partial k}{\partial x_j}\right] = -\bar{\rho}\tau_{ij}.\widetilde{S_{ij}} \; - \; C_\epsilon\frac{\bar{\rho}k^{3/2}}{\Delta}, \tag{5}$$

where $\Delta$ is the mesh size and the default model coefficients are $C_k = 0.094$, $C_\epsilon = 1.048$ [35].

## 2.4 Numerical Method and Validation

An in-house finite-volume solver is used to carry out LESs. This solver is suitable for simulations of turbulent compressible flows of the LES type. A skew-symmetric form of a second-order collocated central differencing scheme with low numerical dispersion and dissipation is implemented in the code, combined with a dissipative semi-discrete, central difference scheme. Hence, it preserves the energy in turbulent regions while capturing any discontinuities in the domain. Switching between the two schemes is performed using an embedded Ducros sensor [36]. This sensor allows the hybrid scheme to capture the discontinuity and predict the accurate decay of turbulent kinetic energy in turbulent regions, making the model suitable for the LES of compressible turbulent flows. The governing equations are time-integrated using the original implicit Euler scheme. The code has been validated for a large number of shock and turbulence interaction cases [37–39] and separated flows [40]. The flow is assumed Newtonian, with a specific heat ratio of $\gamma = 1.4$ and a Prandtl number of $Pr = 0.72$. A detailed comparison of the present LES and experimental results for the reattaching free shear layer, including the reattachment process and downstream boundary layer growth, was previously published [41]. To validate the non-adiabatic wall condition, a careful assessment is conducted, and the results are presented in previous work [25]. A good agreement between the experimental measurements and the LES profile at the corresponding location provided confirmation of the present approach's quality with non-adiabatic wall conditions.

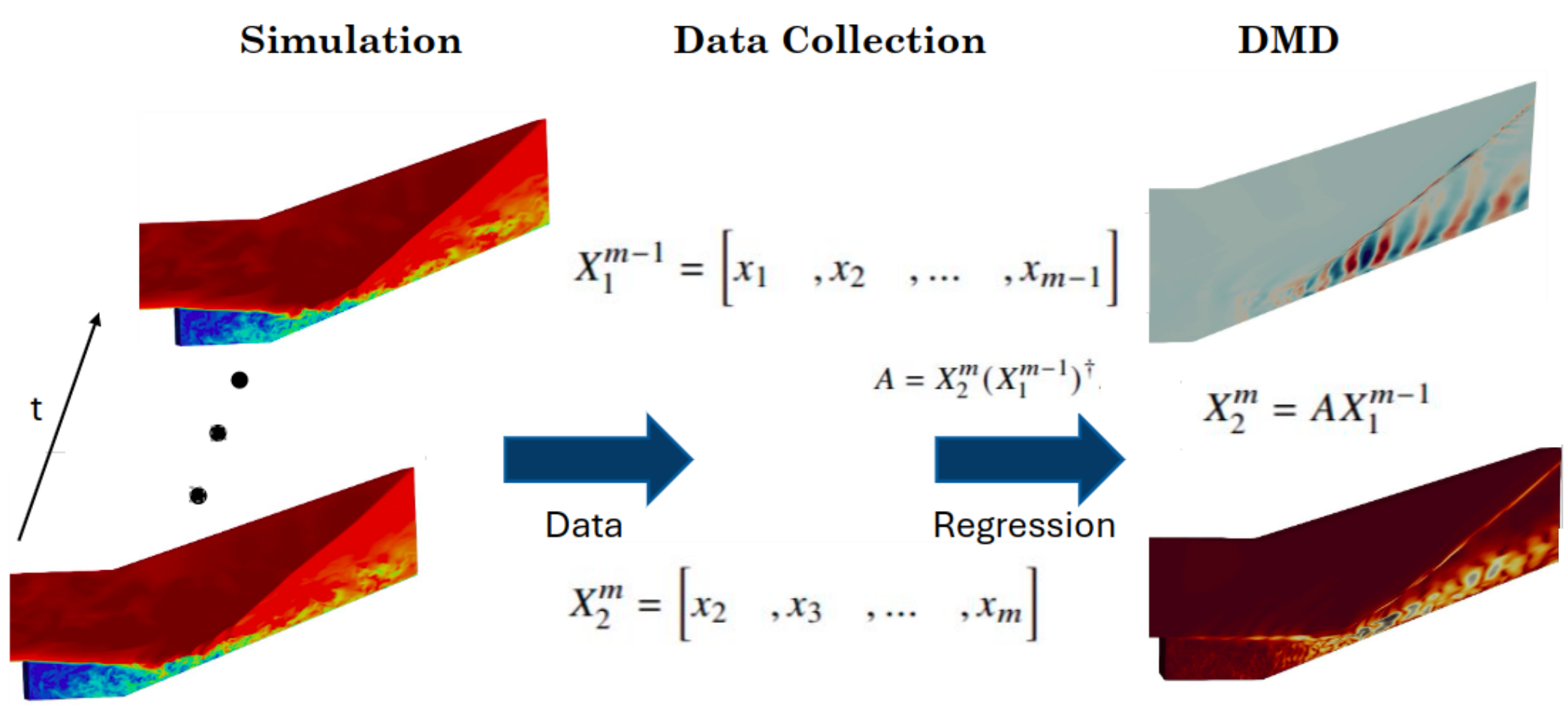

**Figure 4. Extracting most energetic modes of flow field using the mode decomposition technique.**

# 3 Dynamic Mode Decomposition

Model decomposition methods, also known as data-driven model reduction, can decompose large sets of data into modes. In this context, spatial-temporal data can be preserved with spatial modes, comprising information about their temporal variables by extracting dominant coherent structures. This enables converting flow field effects to the most dominant phenomena in a reduced order model (ROM) fashion, which can be analyzed independently. The fundamental idea behind DMD is that any time step of an unsteady flow field can be approximated using a linear combination of previous time steps with a sufficient number of time steps:

$$x_m \approx a_1 x_1 + a_2 x_2 + \ldots + a_{m-1} x_{m-1}. \tag{6}$$

Given the above equation, a linear operator $A$ may be defined such that it can map many consecutive snapshots to their respective snapshot as follows.

$$X_2^m = A X_1^{m-1}. \tag{7}$$

Therefore, the temporal dynamics of the flow field can now be analyzed using eigenvalue of analysis of $A$. Many different ways of evaluating the properties of A have been proposed. One is the Streaming Total DMD (STDMD) algorithm, which is based on the methods shown in Hemati et al.'s work [42]. The core of the algorithm is similar to the process in computing incremental POD modes and consists of the incremental computation of an orthonormal basis, but using the classical Gram- Schmidt iteration (CGSI) process for the generation of an orthonormal basis from augmented snapshots. The detail of the DMD approach is described in Appendix A.

The approach enables the computation of selective modes based on amplitude-related selection criteria, as ordering modes by their physical dominance is crucial for extracting dominant structures. We learned that keeping the first 140 modes captures 99.9% of the total energy and is sufficient for reconstructing the flow field, closely aligning with the results from LES. This is noted as the necessary number of modes for high-dimensional and nonlinear systems [43]. Figure 4 plots the diagram of the DMD method for data collection from simulation data.

# 4 RESULTS AND DISCUSSION

## 4.1 Effect of Wall Temperature on Flow Unsteadiness

We begin with an overview of the flow behavior and qualitative perception of the influence of the wall heat transfer near the mean reattachment point. First, to visualize flow organization in the interaction region, the iso-surfaces of the Q-criterion colored by the instantaneous velocity in the x-direction for the adiabatic wall condition are shown in Fig. 5. The concept of Q-criterion was first introduced by Jeong and Hussain [44] to visualize a vortex, and it is defined as $Q = \frac{1}{2}(||\Omega||^2 - ||S||^2)$ where $\Omega$ and $S$ are rotation and strain rate tensors, respectively. The Kelvin–Helmholtz instability at the edge of the step breaks the upcoming longitudinal structures, resulting in the formation of vortices that are uniform in the spanwise direction. These spanwise structures break down, forming hairpin vortices, as seen in Fig. 5. Near the reattaching point where the shock interacts with the shear layer, hairpin vortices are stretched by the acceleration downstream. Further downstream, these structures are stacked together in the streamwise direction, analogous to the large-scale structures noticed by Ganapathisubramani et al. [45]. The flow is highly unstable as it interacts with the reattaching free shear layer, and turbulence activity is clearly modified behind the shock.

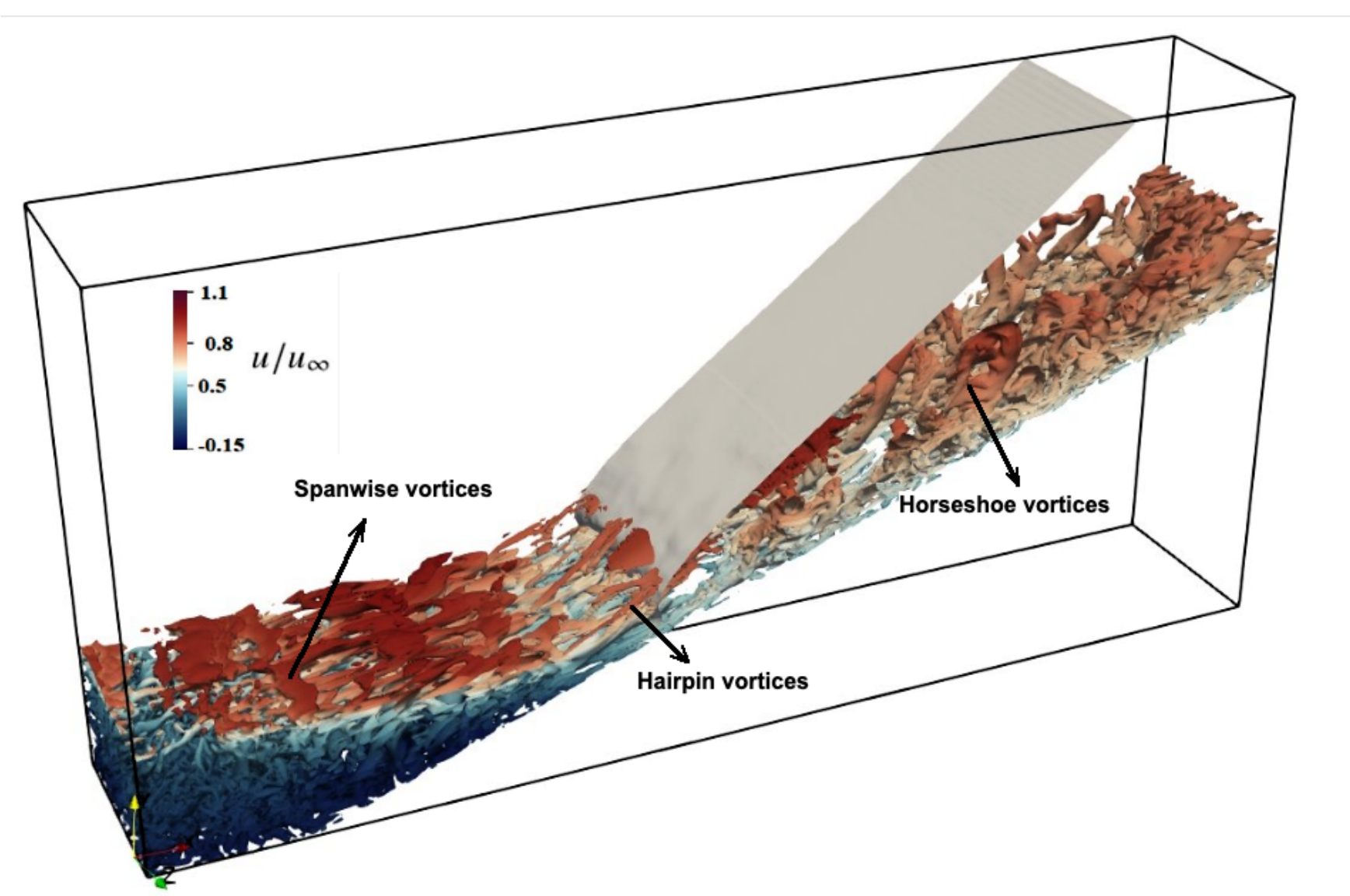

**Figure 5. Iso-surface of the normalized Q criterion colored by the streamwise velocity for the simulation with adiabatic wall condition. The shock is visualized by the pressure iso-surface (gray).**

In the previous work, we carefully assessed the effect of the wall temperature on the position of the mean reattachment point [25]. We showed that the mean reattachment position moves upstream of the slope as the wall temperature decreases. A strong amplification of the heat transfer rate is observed in the interaction region for extreme cases with respect to the reference cooled or heated boundary layers, with a maximum increase of approximately a factor of 1.7 for the cooled wall and 1.5 for the heated wall. These trends are very similar to those reported by Bernardini et al. [23], which explores the effect of temperature on the interaction between supersonic shock waves and TBL. It should be remarked that the reduction of the length scale in the streamwise and wall-normal directions by wall cooling produces stronger temperature gradients at the wall. Correspondingly, since the shock penetrates more deeply into the reattaching shear layer and the pressure jump conveyed by the shockwave must be sustained in a smaller region, cooling the wall increases the root-mean-square wall pressure. The position of the maximum values for pressure superbly matches that of the first peak in the Stanton distribution, implying that the generation of high thermal loads is likely to be linked to the pressure jump in the interaction region.

Figure 6 (a) depicts contours of the instantaneous heat transfer coefficient, Stanton numbers, at the wall plane for the two extreme cases downstream of the interaction as shown in the previous work [25]. We discussed that these 'thermal streaks' are the result of intense wall-temperature fluctuations and can significantly affect heat transfer in the boundary layer [46]. The wall temperature has a significant impact on the initiation and development of such streaks [47]. Figure 6 (b) shows the time average of heat transfer at the wall plane for both cases. These wall planes feature organized longitudinal structures. These structures are referred to as Görtler vortices in the literature [48, 49]. It is established that these vortices are responsible for heat and load amplifications near the reattachment point of separated flows [50]. In particular, it has been shown that nonlinear development of the counter-rotating vortex pair associated with Görtler modes gives rise to bell-shaped structures along the boundary layer [51]. The pair of counter-rotating vortices lifts low-speed fluid from the wall while simultaneously pushing high-speed fluid downward towards the wall on both sides. This interaction causes the boundary layer to be compressed towards the wall between the two neighboring vortex pairs, leading to a significant temperature gradient and creating a hot region on the wall [52]. As a result, striation patterns in heat transfer are observed, along with the distinctive footprints of Görtler-type vortices at the point of reattachment. The evaluation of the effect of wall-temperature conditions on the organization of these Görtler vortices reveals an interesting point. The size and number of these structures depend on the wall temperature condition. Heating the wall surface splits the vortices into slimmer vortices. This phenomenon can explain the higher load and heat transfer associated with the cold wall. The high-fidelity simulations underline significant physics of various wall thermal conditions *per se*. However, the fact that the development, growth, and size of Görtler vortices are linked to wall loading and heat transfer is evidence of strong intercoupling between momentum and heat in a reattaching TBL. Hence, this coupling should be examined using a decomposition approach to isolate the active modes affecting flow dynamics under various heat transfer conditions.

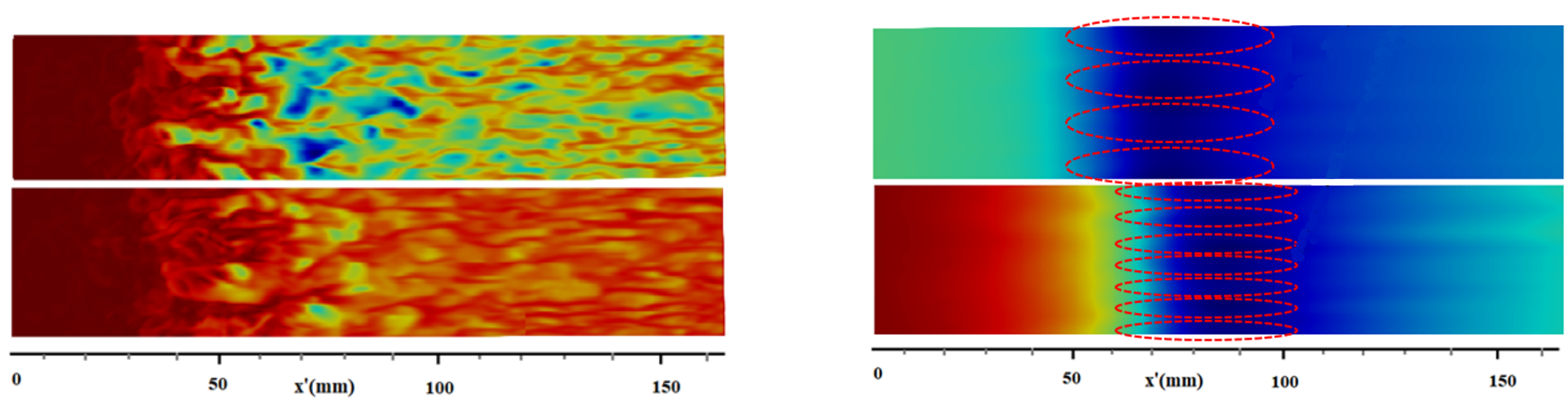

(a) (b)

**Figure 6. a) Contours of instantaneous Stanton number at the wall plane [25] and b) contours of mean Stanton number (Top is the cold wall and bottom is the hot wall).**

## 4.2 DMD of Velocity Field

We deployed the DMD method to the mean velocity fields in the streamwise direction and at the mid-span of the domain for all three cases. A total of $N = 5000$ time snapshots are used with intervals of $\Delta T = 0.01 L_R/U_\infty$ are used where $L_R$ is the separation length on the ramp. To determine the required number of flow snapshots, we checked the convergence of the eigenvalues corresponding to the DMD modes. By calculating the relative error of the L2 norm of eigenvalue spectrum (following Ref. [53]) we realized by increasing the number of snapshots ($M = 1250, 3000, 5000, 6500$) the error gradually decreases as the number of snapshot increases and remains a small value at $M = 5000$, namely, the error does not change by increasing beyond this number of snapshots. Figure 7 shows the polar plots of the computed amplitude (left) and eigenvalues vs. frequency (right) for cold, adiabatic, and hot walls, from top to bottom. From the polar plots, modes appear in complex conjugate pairs, and the one with the largest amplitudes $(\alpha_i \lambda_i)^{m-1}$ is contoured with yellow color. The region of large amplitude extends during heating and decreases slightly during cooling, as shown in the polar plots. Most modes are either stable or slightly damped (i.e., with $|\lambda| < 1$), but there is a tendency to slightly higher damping values in the higher-frequency range. The frequency is nondimensionalized using the separation length, $L_R$. Four peaks can be educed from the amplitude plot for the reference adiabatic case. The first peak may be identified with the lowest frequency,$f_1$, at $St_R \approx 0.16$, another peak, $f_2$, is located at $St_R \approx 0.3$, third peak $f_3$, is located at $St_R \approx 0.33$, and finally, a peak with the highest frequency, $f_4$, is at $St_R \approx 0.4$. It is established that separated-attached flows associate with two frequency modes, which are the consequence of flapping and shedding phenomena of the shear layer [54]. The flapping motion is manifested in the breathing motion of the separation bubble, where the upstream motion of the separation point is linked to the downstream motion of reattachment and vice versa [55]. The mechanism of the elongation and shrinkage of the separated region, i.e, breathing of the separation bubble, is the dominant phenomenon, inferred as a dynamical system that has the lowest frequency (corresponding to the flapping modes), at $St_R \approx 0.16$, as also reported in the literature (Cherry et al. [54] and Kiya et al. [56]). The vortex shedding due to the large scale motion of the mixing layer is associated with the second lowest frequency at $St_R \approx 0.3$ (corresponding to the shedding mode), as also reported by others (Kiya et al. [57] and Tenaud et al. [58]). It can be deduced that the third frequency at $St_R \approx 0.33$ is likely to be a harmonic of the flapping frequency $f_3 \approx 2f_1$. Lastly, in the higher-frequency region, one peak is identified at $St_R \approx 0.4$. This peak is attributed to the Kelvin-Helmholtz instability mode of the separation edge. This is also confirmed by POD analysis of shear layers, which showed that the highest frequency peak is associated with Kelvin-Helmholtz instability, where Delville [59] detected three peaks and the highest frequency in the separation region. This peak is attributed to the Kelvin-Helmholtz instability mode of the separation edge and the Strouhal number for Kelvin-Helmholtz is recorded from many experiments (Bernal et al.[60]).

The top plots in Fig. 7 represent the amplitude of various modes for the cooled surface. The flapping frequency can be identified similarly to the adiabatic surface, yet with a slightly lower amplitude. The second peak, $f_2$, can be identified at $St_R \approx 0.22$, which corresponds to the shedding mode. Compared to the adiabatic wall, the shedding mode not only has a higher amplitude, but it also has a lower frequency. The harmonic of the flapping frequency is also apparent at $St_R \approx 0.32$. The Kelvin-Helmholtz instability mode,$f_4$, is similar to the adiabatic surface, which may be detected at $St_R \approx 0.4$. The bottom plots exhibit the hot surface frequency modes. A frequency band is detected for the flapping mode at $St_R \approx 0.15$ with a higher amplitude compared to the adiabatic case. The shedding frequency $f_2$ does not change significantly compared to the reference adiabatic wall. On the other hand, the Kelvin-Helmholtz instability mode has a slightly lower frequency for the heated wall ($St_R \approx 0.38$).

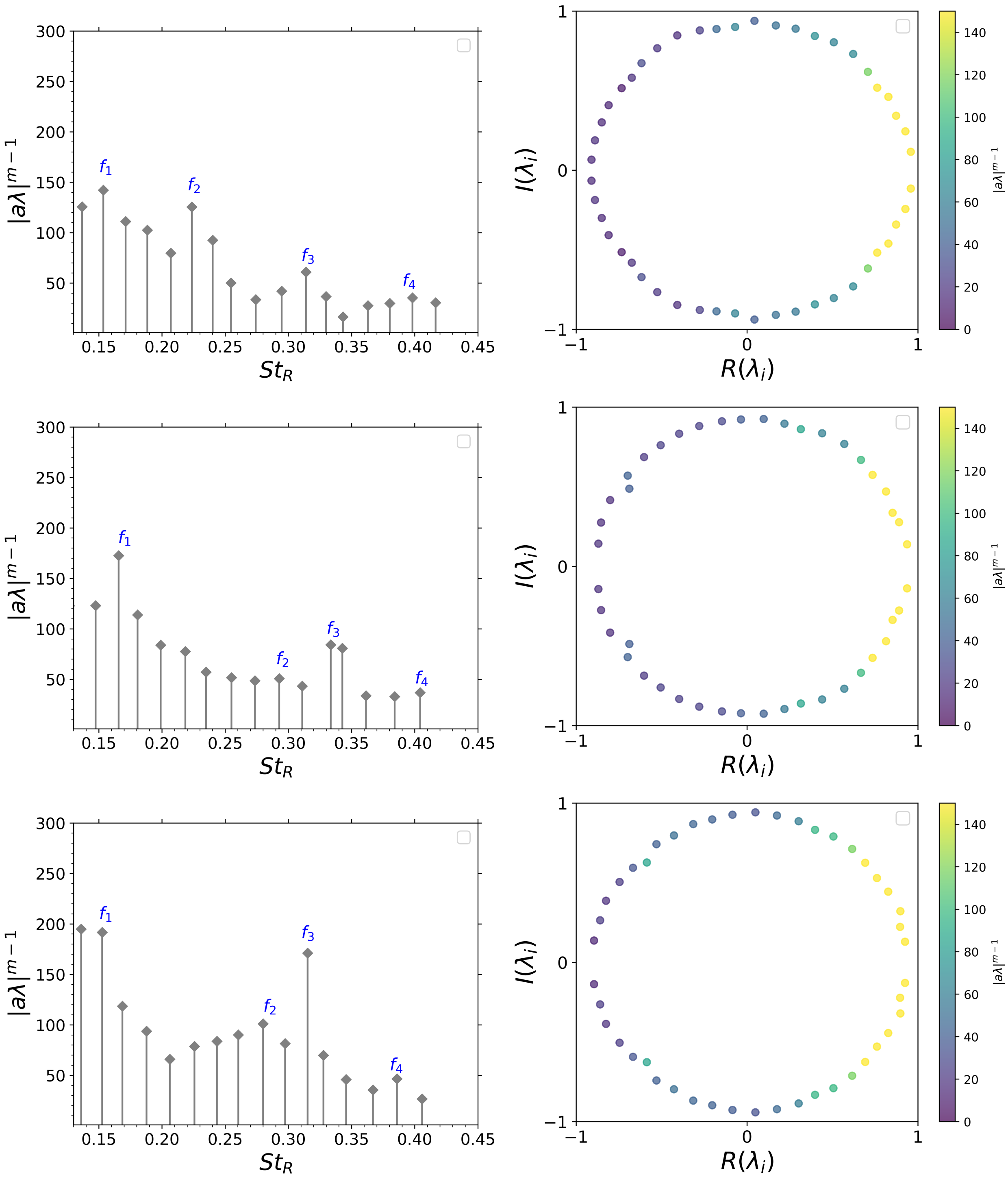

(a) (b)

**Figure 7. a) Amplitude distribution of modes vs frequency calculated from the DMD and b) polar plot of eigenvalues of various modes for cold, adiabatic, and hot walls from top to bottom.**

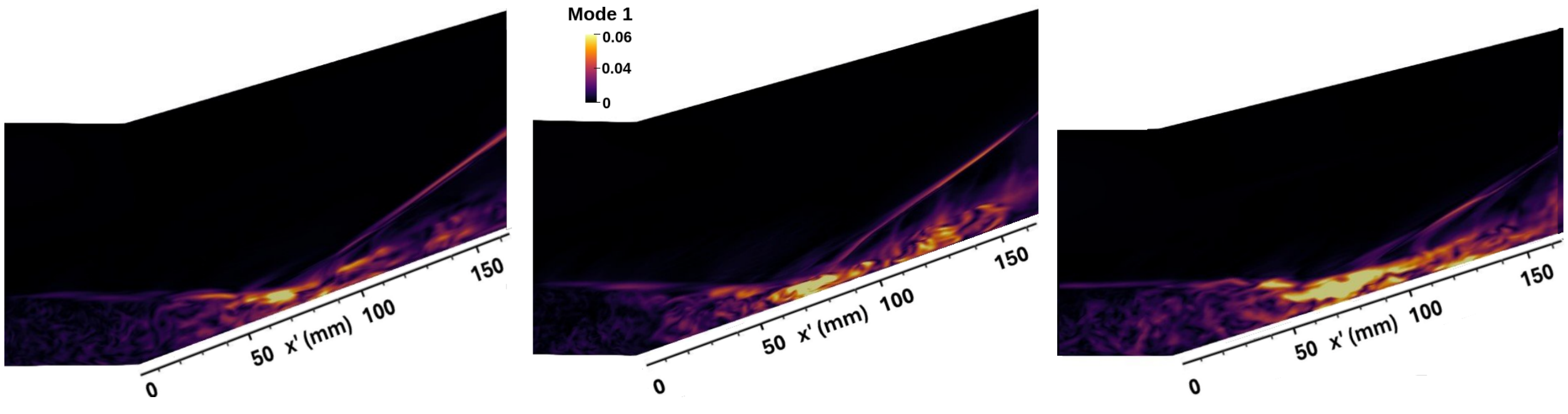

(a) (b) (c)

**Figure 8. Contours plot of shear layer flapping mode for a) cooled wall, b) adiabatic wall and c) heated wall**

Figure 8 reveals the contours of the flapping mode for three wall conditions near the reattachment. The vortical motions are most intense in the vicinity of the reattachment point, where hairpin vortices are stretched by the acceleration downstream. It is evident that, from left to right, as the surface temperature increases, the wall experiences a more intense flapping mode; as shown in Fig. 7, the mode amplitude increases with wall temperature. A frequency band is detected for the hot wall. The flapping mode is linked to the extension and shrinkage of the bubble size. It has been shown that the size of the separation bubble increases with heating [25]; therefore, this may explain the more energetic flapping mode in this case. Figure 9 shows the contour of the shedding frequency for the cold (left) and hot (right) walls. As we already pointed out from Fig. 7, the shedding frequency of the cold wall is lower with a higher amplitude compared to the hot wall. In order to obtain a deeper insight into the shedding mode, one may nondimensionalize this frequency using the height of the step equivalent to the height of the bubble in the separation zone. For the adiabatic wall, one would get $St_H \approx 0.12$. which corresponds to the Karman instability, as stated by Roshko [61]. In the Karman instability mechanism, the interaction of the bubble with the wall generates a row of aligned vortices of opposite sign. These vortices are created by reflection and convect outside the recirculation bubble. It can be deduced that cooling the wall surface results in a lower frequency shedding mode and therefore generates larger reflected vortices, as shown in Fig. 6 (b), those Görtler vortices of the cold wall compared to the slimmer vortices of the hot surface. The intense shedding mode of the cooled surface is apparent from Fig. 9. Figure 10 features the Kelvin-Helmholtz instability mode for two wall temperature conditions. This mode exhibits high-frequency oscillations because it is related to turbulence structures in the upcoming shear layer resulting from a corner break. Comparing the contours of this mode with the shedding mode in Fig. 9 remarks that this mode is most likely coupled with the shedding mode since the pattern is similar. This could be due to the interaction between the high-frequency motions and the wall-shedding vortices.

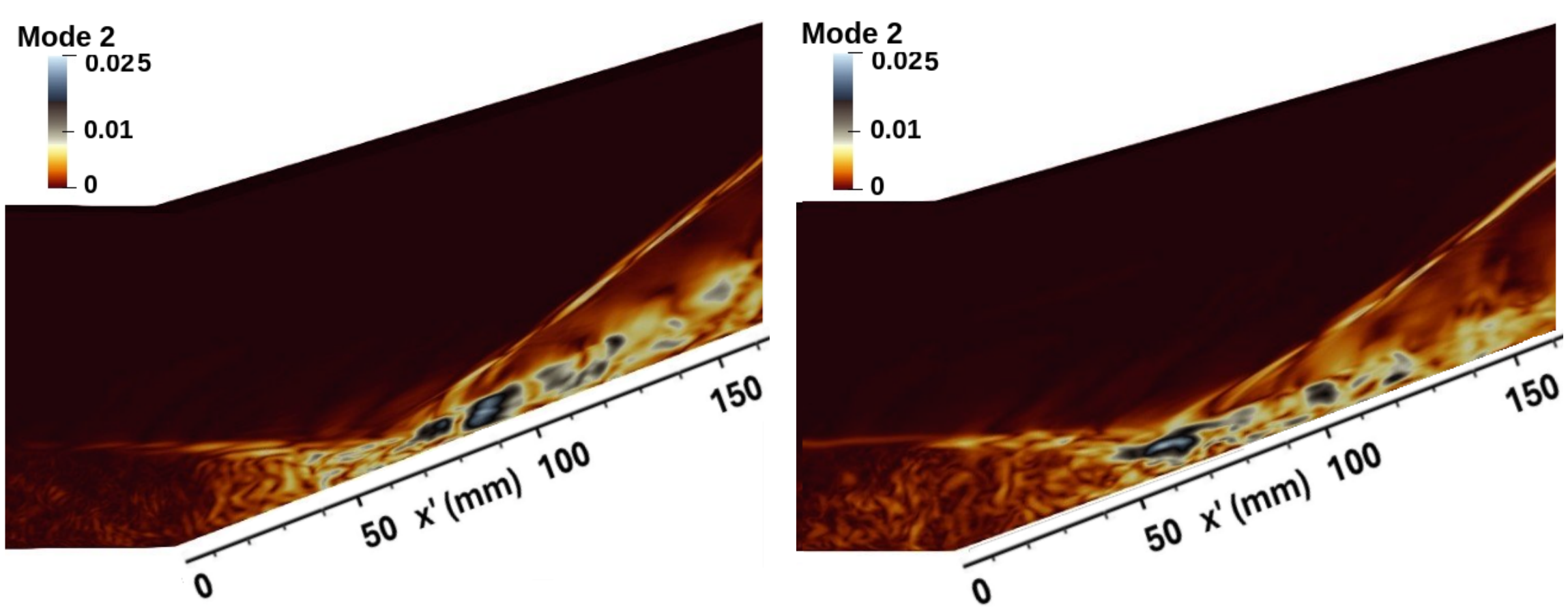

(a) (b)

**Figure 9. Contours plot of shear layer vortex shedding mode in the domain for a) cooled wall and b) heated wall.**

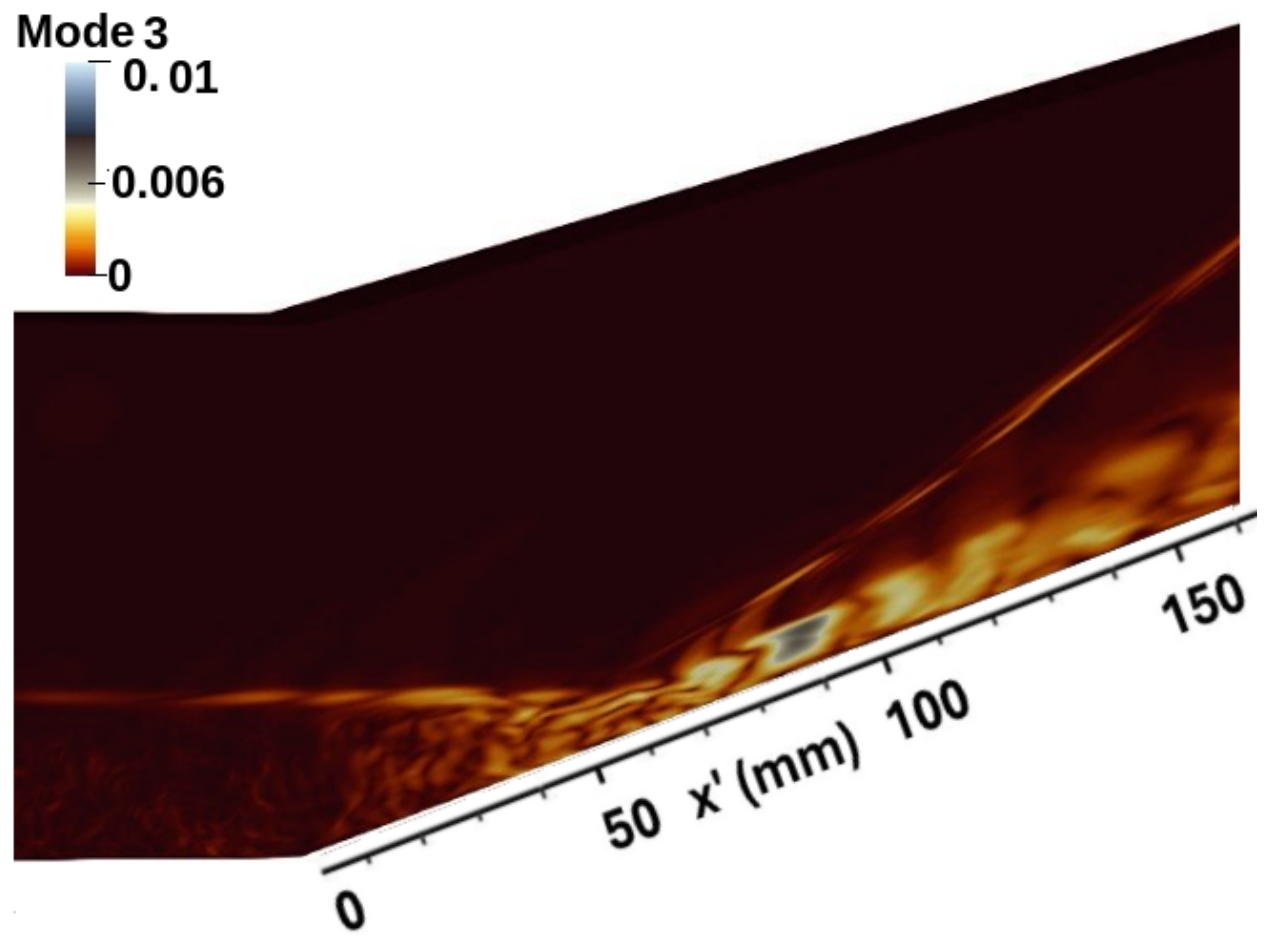

(a)

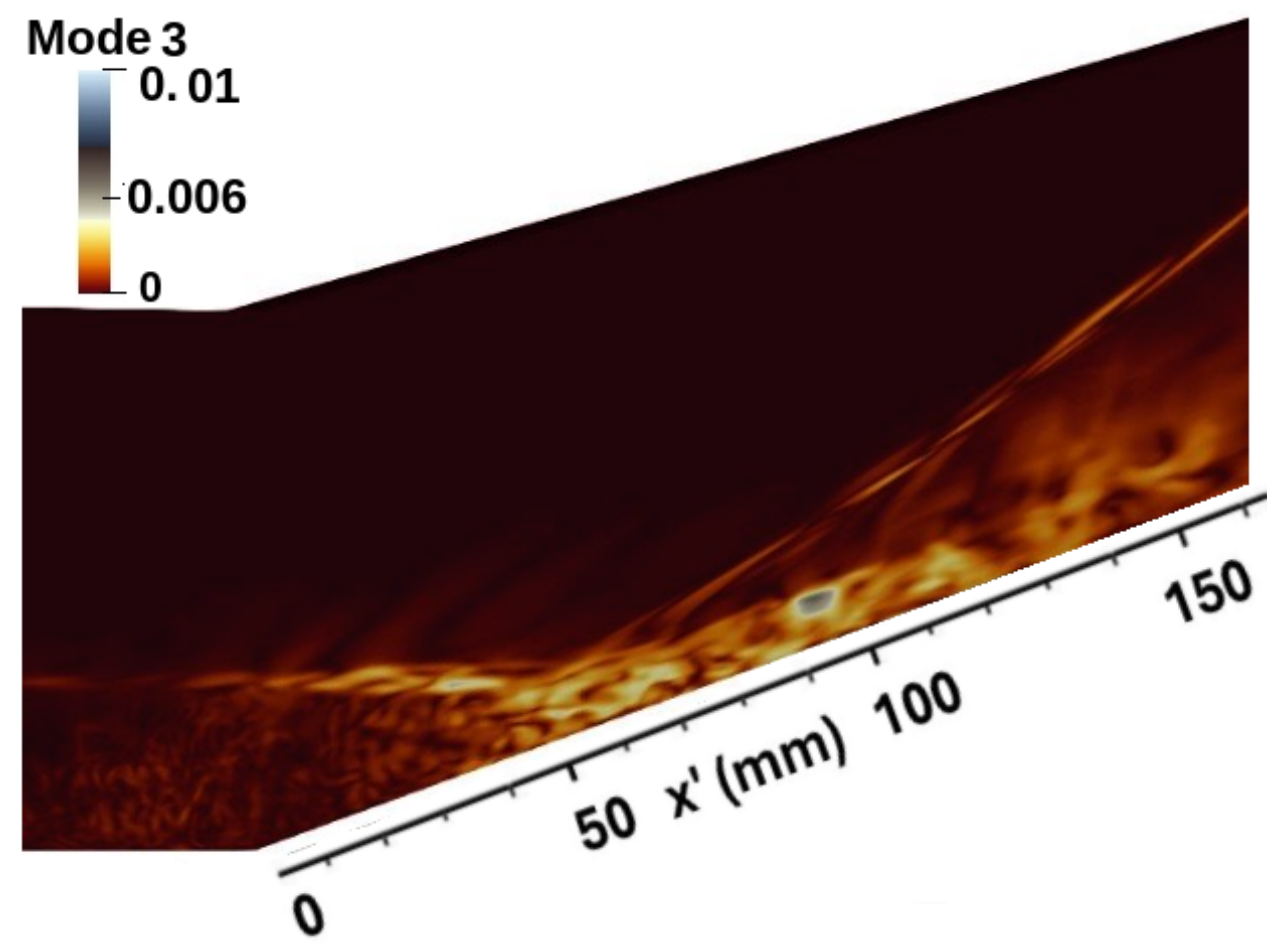

(b)

**Figure 10. Contours plot of Kelvin-Helmholtz instability mode in the domain for a) cooled wall and b) heated wall.**

Figure 11vs. frequency $f$, where $p'$ is the pressure fluctuations and $\tau = L_R/u_\infty$ is the time constant, which is defined using the length of the separation bubble on the ramp, $L_R$. It is evident near the reattachment point. A higher magnitude of spanwise vorticity values is apparent for the cooled wall. On the contrary, the hot surface exhibits a less intense vortex footprint near the reattachment point with a substantially smaller magnitude of spanwise vorticity. Note that the reattachment location moves downstream by heating the wall; therefore, the trace of vortices is located further to the right for the hot wall. This observation is of particular interest since comparing it to the contour of mean heat flux profiles in Fig. 6 reveals the strong interconnection between the momentum and heat transfer near any separation zone. The underlying reason for such coupling is linked to the shedding mode of the separation bubble, as discussed before. This interconnection manifests itself in the contours of mean heat fluxes and velocity streamlines.

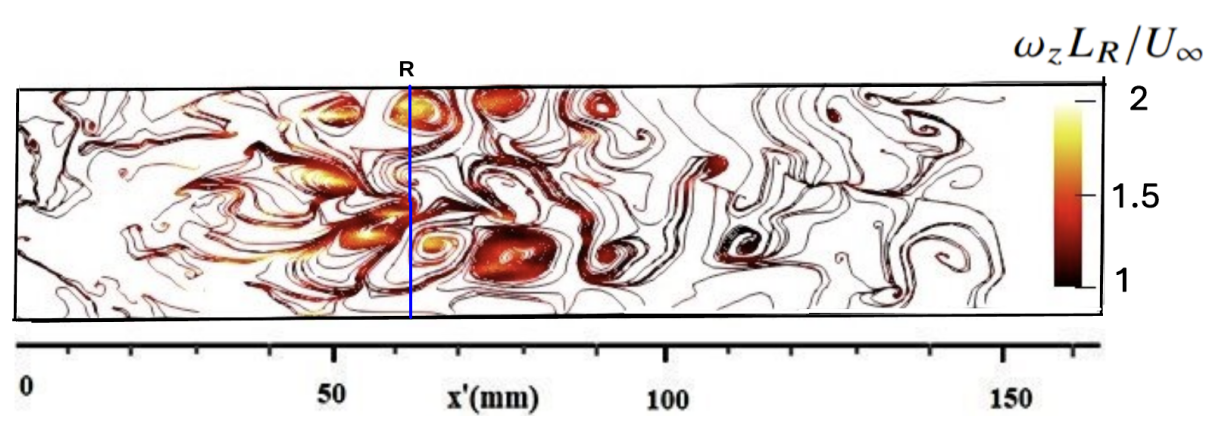

(a)

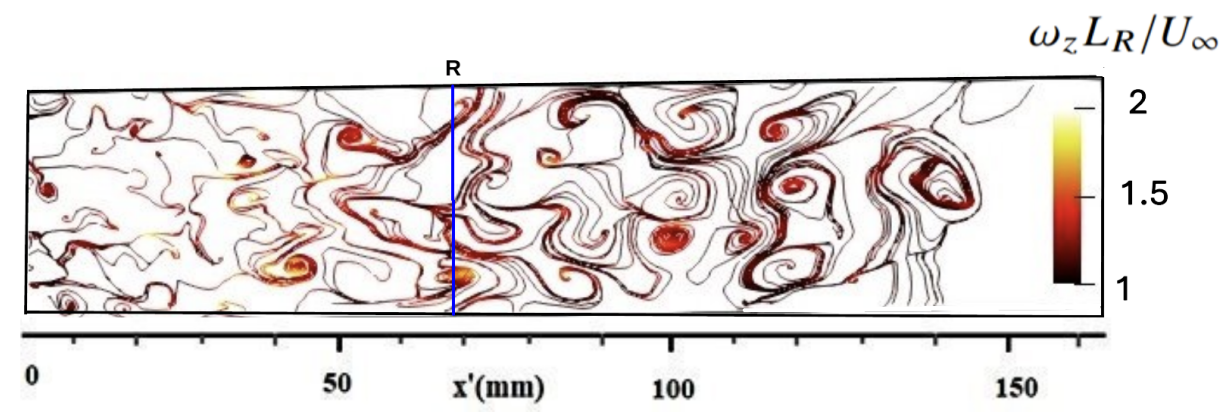

(b)

**Figure 11. Velocity streamlines on the wall plane and near reattachment point contoured by spanwise vorticity for a) cooled wall and b) heated wall.The reattachment point is mark with letter R.**

## 4.3 Surface Temperature and Pressure Fluctuation

The primary mechanism for the shock oscillation in the reattaching shear layer configuration is believed to be perturbations by disturbances in the incoming turbulent flow [9, 62]. The main contributing factors to this unsteadiness are shear layer flapping, breathing of the separation bubble, and vortex shedding from the shear layer [63]. The coupling may occur between all three factors or between any two of them, as shown in the previous section: the coupling between the shedding mode and the Kelvin-Helmholtz instability. A record of fluctuating wall pressure is taken on the reattachment ramp to evaluate the effects of various wall temperatures on shock unsteadiness. Figure. 12 shows the corresponding auto-spectra of the fluctuating pressure signal, $G(f) = \frac{4\overline{p'}^2\tau}{1+(2\pi f\tau)^2}$ vs. frequency $f$, where $p'$ is the pressure fluctuations and $\tau = L_R/u_\infty$ is the time constant, which is defined using the length of the separation bubble on the ramp, $L_R$. The spectra profiles are plotted for two locations, $x = x_R$ and $x = 1.05x_R$. The data are seen to have a broadband energy content with no prominent peaks.The higher frequencies correspond to the eddies present in the separation bubble. As marked in Fig. 12 (b), the energy cascades, i.e., transfer from large scale to small scale flow motions follows $k^{-5/3}$ marginally.

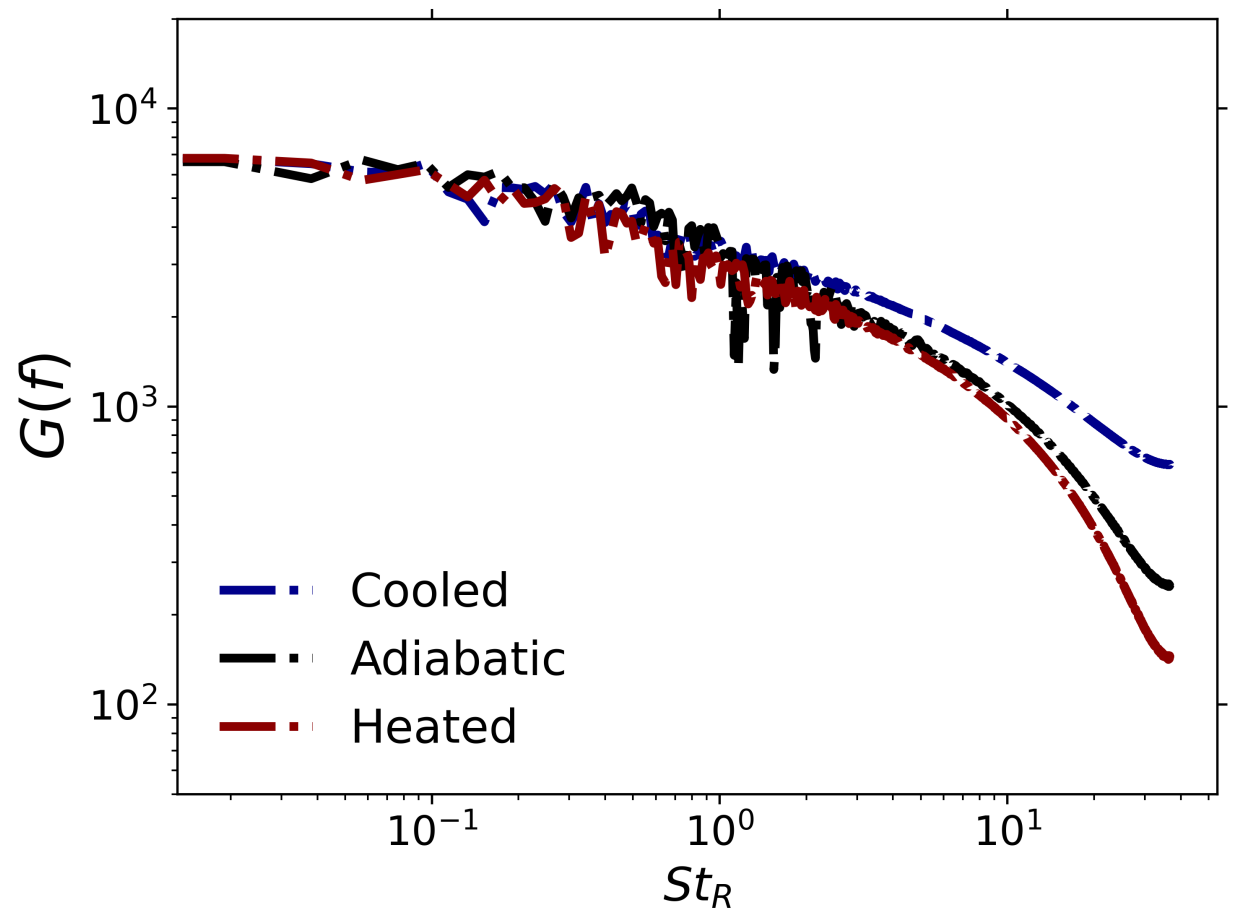

(a)

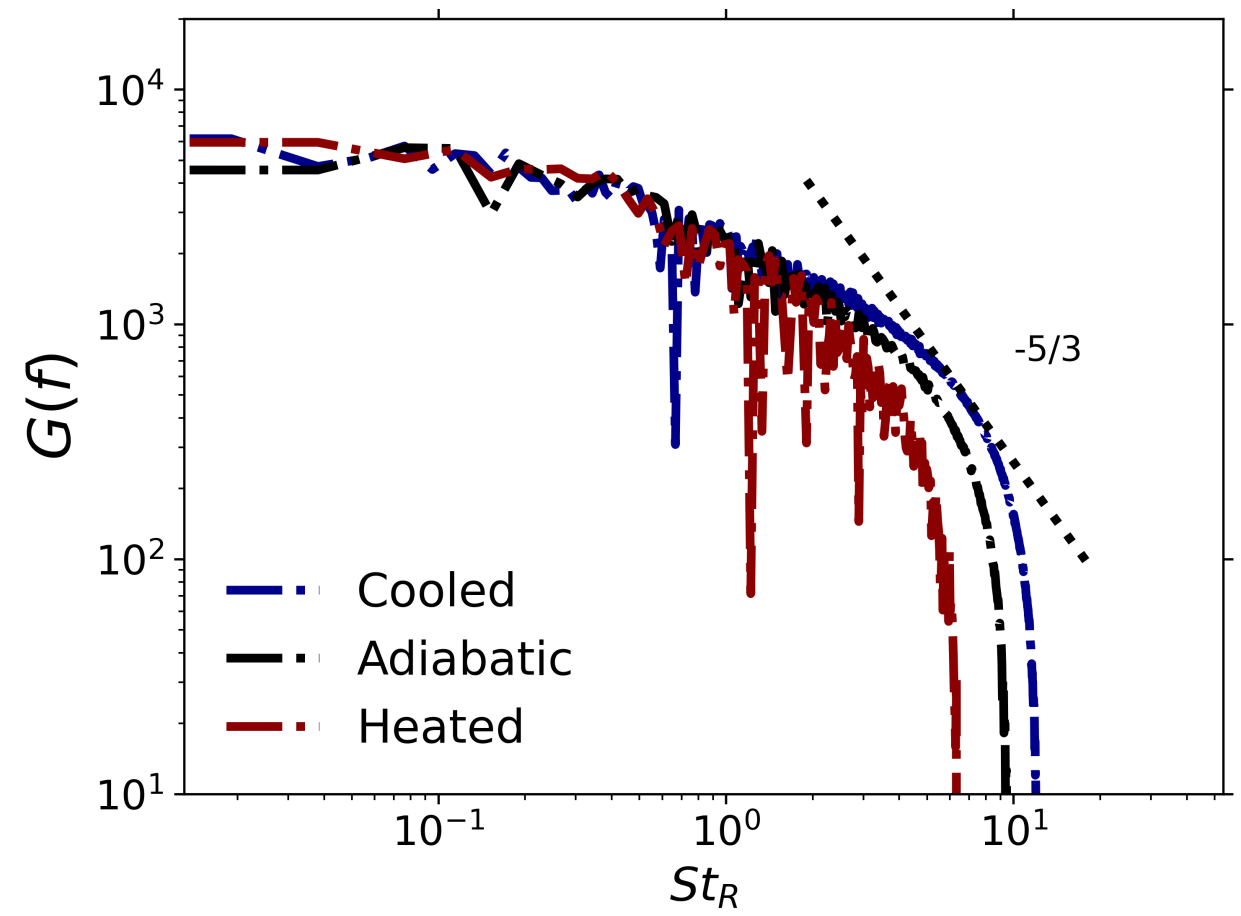

(b)

**Figure 12. Auto-spectra of the wall pressure fluctuation data on the ramp at a)** $x = x_R$**, b)** $x = 1.05x_R$**.**

Comparing the energy spectra at two locations shows that reducing the energy content at lower frequencies occurs downstream of the mean reattachment point ($x = 1.05x_R$) in the redeveloping boundary layer. It is apparent that the energy content at higher frequencies is superior at $x = x_R$ due to the vortices in the separation bubble. At the mean reattachment point, $x = x_R$ shown in Fig. 12 (a), heating the wall is seen to shift the spectrum down in the higher frequencies region, reflecting a decrease in the intensity of pressure fluctuations, and moves it up with cooling the wall. This is aligned with the observation from the previous section. The more intense shedding mode of the cold wall results in the higher energy spectra, most likely due to the reflecting vortices from the wall. Analogous features in the spectra are also seen to be shifted to the right in the heated wall case, reflecting an increase in characteristic frequency. Namely, the high-frequency motion due to turbulence in the separation bubble becomes dominant by heating the wall, as it is apparent in Fig. 12 (b).

# 5 CONCLUSION

It is reported that wall heating causes the reattachment location to shift downstream in the reattachment turbulent shear layer. Conversely, wall cooling moves it upstream on the ramp but has the drawback of increasing the wall pressure fluctuations. The distribution of Stanton numbers indicates that the cooling wall has a strong heat transfer. The mean heat transfer contours reveal the presence of organized Gottler vortices downstream of the reattachment point. The size and number of these structures depend on the wall temperature condition. Heating the wall surface splits the vortices into slimmer vortices. To explore the underlying mechanisms that drive the momentum and heat interconnections, we applied a mode decomposition method to the mean velocity field to extract a linear system. From the DMD results, following facts can be drawn from the DMD results:

- The shear layer flapping is the most energetic mode at $St \approx 0.16$. The hot wall gains the highest amplitude at the flapping frequency and the vortical motions are most intense in the vicinity of the reattachment point for the heated wall.
- The vortex shedding due to the large scale motion of the shear layer is associated with the second lowest frequency. The cold wall not only has a higher amplitude of the shedding mode, but it also has a lower frequency compared to the adiabatic wall.
- The highest frequency peak, which is associated with the Kelvin-Helmholtz instability mode, is found to be coupled with the shedding frequency.

- The auto-spectra of wall fluctuating pressure reveals that he more intense shedding mode of the cold wall results in the higher energy spectra, most likely due to the reflecting vortices from the wall.
- The streaks observed in the heat transfer contour of the cooled wall case result in an intense temperature gradient near the wall. These streaks are footprints of vortices from the shedding mode of the shear layer, which is shown to be more energetic when the wall surface is cooled.

This work sheds light on the underlying physics for the intercoupling of momentum and heat, with the goal of mitigating aircraft fatigue loading caused by intense wall pressure fluctuations and heat flux. However, our study is limited to one Reynolds and Mach number. Future investigations should focus on the effect of various parameters, such as the Reynolds number and Mach number. Practical applications require a knowledge of the effects of Mach number, Reynolds number, wall to free-stream temperature ratio and wall geometry, especially curvature, on the interconnection of heat-transfer and momentum of the separated and reattaching boundary layer.

# Appendix A

STDMD employs a QR- decomposition of the augmented snapshot matrix $Z$ as

$$Z = Q_Z R_Z. \tag{8}$$

Matrix $X_1^{m-1}$ is used as the first part of the snapshot matrix containing the first $m-1$ snapshots, and $X_2^m$ contains the last $m-1$ snapshots. The upper part of $Z$ can also be represented as

$$X_1^{m-1} = \begin{bmatrix} 1 & 0 \end{bmatrix} Q_z R_z. \tag{9}$$

$$(X_1^{m-1})^T = Q_z^T R_z^T \begin{bmatrix} 1 \\ 0 \end{bmatrix}. \tag{10}$$

and the lower part as

$$X_2^m = \begin{bmatrix} 0 & 1 \end{bmatrix} Q_z R_z. \tag{11}$$

Because $R_z$ is not explicitly required for the evaluation of DMD modes, only

$$G_z = R_z R_z^T \tag{12}$$

is evaluated during the incremental updates. The expression for the evaluation of the DMD operator is the starting point for the derivation

$$A = X_2^m (X_1^{m-1})^\dagger, \tag{13}$$

where † is the pseudoinverse. Using the identity

$$X^\dagger = X^T (X_1^{m-1} (X_1^{m-1})^T)^{-1}, \tag{14}$$

equation

$$A = X_2^m (X_1^{m-1})^T (X_1^{m-1} (X_1^{m-1})^T)^{-1} \tag{15}$$

is obtained. To arrive at a valid definition of an STDMD operator, the definition of another QR-decomposition

$$Q_x R_x = \begin{bmatrix} 1 & 0 \end{bmatrix} Q_z, \tag{16}$$

is needed. $Q_X$ is a more noise-insensitive representation of the image of the snapshot matrix. It is worth noting that the above QR-decomposition should not be mistaken for the QR-decomposition of snapshot matrix $X_1^{m-1}$. Rearranging Eq. 9 for $Q_z$ and inserting into Eq. 16 leads to

$$Q_X R_X = X_1^{m-1} R_z^{-1}. \tag{17}$$

After simplification and rearrangement for the snapshot matrix, $X_1^{m-1}$,

$$X_1^{m-1} = Q_X R_X R_z, \tag{18}$$

$$(X_1^{m-1})^T = Q_X^T R_X^T R_z^T \tag{19}$$

is obtained. Finally, the STDMD operator is found by substituting all matrices on the right-hand side of Eq. 15 with expressions (10), (11), (18), and (19) as

$$A = \begin{bmatrix} 1 & 0 \end{bmatrix} Q_z R_z R_z^T Q_z^T \begin{bmatrix} 1 \\ 0 \end{bmatrix} (Q_x R_x R_z^T R_z R_x^T Q_x^T)^{-1}. \tag{20}$$

Another definition is the combination of terms on the right-hand side using definition in Eq. 12

$$R_x R_z R_z^T R_x^T = R_x G_z R_x^T = G_x, \tag{21}$$

After simplification, the projected STDMD operator is then given as its projection onto the orthonormal basis $Q_x$ with

$$A_p = Q_x^T A Q_x = Q_x^T \begin{bmatrix} 1 & 0 \end{bmatrix} Q_z G_z Q_z^T \begin{bmatrix} 1 \\ 0 \end{bmatrix} Q_x G_x^{-1}. \tag{22}$$

As is the case with other variants of DMD, projecting the DMD operator onto the orthonormal basis representation of the snapshot matrix turns out to be computationally more manageable due to the lower dimensionality of the orthonormal basis in case of singular value truncation, or online orthonormal basis compression respectively. The STDMD complex number eigenvectors and eigenvalues are computed using the eigen decomposition of $A_p$, the eigenvectors $y_k$ are evaluated by solving the eigenvalue problem

$$A_p y_y = \lambda_k y_k, \quad A_p Y = \Lambda Y. \tag{23}$$

For the evaluation of the complex-valued STDMD modes, the eigenvectors of $A_p$ are then applied onto the orthonormal basis $Q_x$

$$\varphi_k = Q_x y_k. \tag{24}$$

Eq. 24 is known from Hemati et al. [64] work on STDMD and produces excellent results. To further reduce the amount of memory required for the evaluation of DMD modes, $Q_x$ may be replaced with rearranged Eq. 16 with

$$Q_x = \begin{bmatrix} 1 & 0 \end{bmatrix} Q_z R_x^{-1}, \tag{25}$$

the following equation for the projected STDMD operator can be obtained

$$A_p = (\begin{bmatrix} 1 & 0 \end{bmatrix} Q_z R_x^{-1})^T \begin{bmatrix} 1 & 0 \end{bmatrix} Q_z G_z Q_z^T \begin{bmatrix} 1 \\ 0 \end{bmatrix} (\begin{bmatrix} 1 & 0 \end{bmatrix} Q_z R_x^{-1}) G_x^{-1}. \tag{26}$$

Using the same replacement for Eq. 24, results in

$$\varphi_k = \begin{bmatrix} 1 & 0 \end{bmatrix} Q_z R_x^{-1} y_k, \tag{27}$$

$$\Phi = \begin{bmatrix} 1 & 0 \end{bmatrix} Q_z R_x^{-1} Y. \tag{28}$$

These formulations in Eqs. 27-28 also have substantial advantages in the evaluation of amplitudes. The frequency of each DMD mode is computed by

$$f = \frac{Im[\log \lambda_k]}{2\pi \Delta t} \tag{29}$$

where $\Delta t$ represents the fixed time step size between snapshot $x_m$ and $x_{m+1}$. The mode amplitudes can be computed as

$$\alpha = Y^{-1}(R_x^{-1})^T Q_z^T \begin{bmatrix} 1 \\ 0 \end{bmatrix} x_1. \qquad (30)$$

## Acknowledgments

This work used STAMPEDE3 at TACC through allocation PHY230146 from the Advanced Cyberinfrastructure Coordination Ecosystem: Services and Support (ACCESS) program, which is supported by U.S. National Science Foundation grants #2138259, #2138286, #2137603, and #2138296.

## AVAILABILITY OF DATA

The data that support the findings of this study are available from the corresponding author upon reasonable request.

## Declaration

**Funding:** This project has no funding.
**Conflict of Interest:** The authors have no conflicts of interest to declare.